\documentclass[prl,twocolumn,floatfix]{revtex4}
\usepackage{graphicx}
\usepackage{amssymb}
\usepackage{dcolumn}
\usepackage{bm}
\begin{document}

\title{Supersonic crack propagation in a class of lattice models of Mode III brittle fracture}
\author{T. M. Guozden}
\author{E. A. Jagla}
\affiliation{Centro At\'omico Bariloche and Instituto Balseiro \\ 
Comisi\'on Nacional de Energ\'{\i}a At\'omica, (8400) Bariloche, Argentina}
\date{\today}

\begin{abstract}

We study a lattice model for mode III 
crack propagation in brittle materials in a stripe
geometry at  constant applied stretching. 
Stiffening of the material at large deformation 
produces supersonic crack propagation.
For large stretching the propagation is guided by 
well developed soliton waves.
For low stretching, the crack-tip velocity 
has a universal dependence on stretching  that can be obtained using a 
simple geometrical argument. 

\end{abstract}

\maketitle

The determination of crack-tip velocities in brittle fracture is one 
of the unsolved problems in fracture mechanics\cite{freund,broberg}. 
Spontaneous crack propagation typically occurs when some energy balance condition 
is satisfied. This is usually known as the ``Griffith's criterion" and it
states roughly speaking, that the available elastic energy 
has to be larger than the energy necessary to create new 
surface in the material upon crack advance. 
We will define the excess elastic energy (EEE)
as the difference between the  available elastic energy and the (static) crack-surface energy.
When the EEE is positive a dynamic regime is achieved. 
The nature of this regime depends on the 
value of the EEE\cite{freund,broberg,fineberg}. If the 
EEE is low, a stationary regime is usually observed, in which the crack 
velocity is a well defined (material dependent) function of the EEE, and if the geometric configuration is symmetric
with respect to the crack, crack propagation occurs along a 
straight line. For larger values of the EEE, crack path oscillations, 
crack branching and other phenomena typically occur. 
 
We will focus 
here in the low EEE regime. Even in this relatively simple case, the
prediction of crack-tip velocity as a function of material parameters, 
sample geometry and strain conditions is a challenging problem. 
Linear elastic fracture mechanics (LEFM)\cite{freund,broberg} suggests
that the crack-tip velocity $V$ should 
be a univocus function of the EEE and material characteristics. Among 
these, microscopic details of the process zone must be included 
in general. Another most important prediction of LEFM is the impossibility 
that the value of $V$ exceeds the Rayleigh velocity of the material, which is typically
somewhat lower than 
the shear sound velocity. 
There is some experimental evidence that challenges this prediction\cite{goma}. In addition, it 
has been recently shown\cite{gao} that elastic 
stiffening at large deformation can unambiguously produce cracks running 
faster than the shear sound velocity in simulations of mode I (opening) propagation. 

Here we analyze the dependence of crack tip velocity upon model parameters in a 
class of piece-wise harmonic lattice model for mode III fracture.
We find supersonic propagation and study its characteristics.
We find a remarkable regime in which the crack-tip velocity can be accurately 
predicted on the basis of a geometric argument, and is independent of most of the
model parameters. 

We model a stripe of material,
laterally stretched in the out-of-plane direction, with a crack
propagating along the mid line. This 
geometry is particularly adequate to study stationary crack propagation
as the system looks identical upon extension of the crack. 
This kind of models have been studied before, for instance in Refs. \cite{slepyan,kessler}.
The model consists of a discrete
square lattice. 
At each node a mass is placed, and the links to nearest neighbor sites are modeled
by piece-wise linear springs. 
The full lattice forms a rectangular system, to which out-of-plane fixed-displacement lateral boundary conditions
are applied. 
We take the horizontal $x$ axis along the stripe. This is also the crack propagation direction.
The out-of-plane displacement
of the masses (the only one which is non-zero) is noted by $u$. The equation of motion 
of a mass at the discrete position 
$i,j$ in the lattice is simply

\begin{equation}
\frac {d^2 u_{i,j}}{dt^2}=\sum_{[i',j']}F(u_{i',j'}-u_{i,j})
\label{eq}
\end{equation}
where $[i',j']$ are the neighbor sites to $i,j$ and
the function $F$ specifies the force law of the corresponding spring:
$F(x)=x$ for $x < u_{nl}$, and $F(x)=\gamma x$ for $x>u_{nl}$ for intact springs, and $F(x)=0$
for broken springs.
Note that the low deformation spring 
constant and mass of the particles have been set to 1 (this sets the wave velocity to 1, also), and that
$u_{nl}$ is the critical stretching at which the spring constant changes from 1 to $\gamma$ ($\gamma >1$).

Springs are assumed to break irreversibly when the stretching is larger than a fixed threshold value
$u_{bk}$. 
To avoid crack branching, which may occur under certain conditions, we allow to break only the 
vertical springs in the center
of the stripe (for low stretching this procedure is not necessary, see below). 
For symmetry reasons we simulate only the upper half of the
stripe, and $N_y$ will denote the number of rows in this half system. 
The boundary condition at the bottom border is
$u_{i,1}=-u_{i,0}$, whereas at the upper border we have $u_{i,N_y+1}=(N_y+1/2)\delta$, where $\delta$ is the
nominal stretching applied to each vertical spring.
We always keep $\delta< u_{nl}$, not to force the system in the non-linear regime with the
uniform applied strain.
Along the $x$ direction, we take the boundary
conditions as follows: $u_{N_x,j}=(j-1/2)\delta$ and $u_{0,j}=(N_y+1/2)\delta$. 
These conditions correspond to those that should 
be valid at $x\rightarrow \pm \infty$. By enforcing them at a finite $x$-position some error is introduced,
but we check that this error is small if $N_x$ is large enough. The crack tip is initially 
placed at an $x$ position close to the right border of the system, by seeding the simulations 
with an appropriate initial condition. 
Along the simulation the crack advances in the positive
$x$ direction. Each time we detect that a new spring ahead of the crack has broken, 
we shift all system coordinates and velocities one unit to the left.
In this way the crack tip is always maintained 
at the same position with respect to the lattice, and the instantaneous crack-tip velocity is 
determined as the inverse of the difference in consecutive times of spring breaking. Reported values correspond to 
averaging over long simulation times, after dependences on initial conditions have died out.

In the present stripe geometry, the Griffith's criterion is stated as follows:
crack propagation cannot occur if stretching is lower than some critical value $\delta_G$. 
Equating the elastic energy far ahead of the
crack-tip, to that far behind it, we find $\delta_G=\sqrt{2\Gamma/(2N_y+1)}$, where
$\Gamma=u_{bk}^2/2+(\gamma-1)(u_{bk}-u_{nl})^2/2 $ is the crack energy per bond. We see that $\delta_G\rightarrow 0$ for
$N_y\rightarrow\infty$.
The EEE per horizontal bond (referred to as $\varepsilon$) is easily calculated in terms of $\delta_G$ as
$\varepsilon=\Gamma\left[ (\delta/\delta_G)^2-1\right]$.
As it should, this quantity vanishes at $\delta=\delta_G$.
Note that contrary to the usual assumption in LEFM, in our model the excess energy $\varepsilon$ 
is not dissipated at the crack
tip, but it is transformed into kinetic energy, that
eventually escapes from the left border of the system.

\begin{figure}
\includegraphics[angle=-90,width=8cm,clip=true]{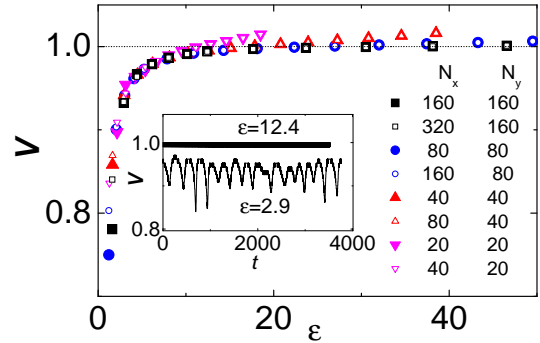}
\caption{(Color online) Average velocity of the crack in the harmonic case 
as a function of the excess elastic energy per horizontal bond $\varepsilon$ 
for different stripe widths $N_y$. To check for independence on stripe length, two different values
of $N_x$ are reported. In the inset, values of the instantaneous velocity as a function of time
for a system of $N_x=320,N_y=160$ are indicated in two cases.
}
\label{f2}
\end{figure}

To test the classical prediction in the normal (sub-sonic) case, 
in Fig. \ref{f2} we show curves for the velocity as a
function of stretching  for $\gamma=1$ and different values of $N_y$. 
The curves converge to a well defined behavior for large $N_y$
if the plot is done 
as a function of $\varepsilon$. 
The result agrees with the predictions of LEFM. The velocity approaches the wave speed for large $\varepsilon$,
whereas at low $\varepsilon$ is reduced due to lattice trapping effects.
In accordance also with
previous simulations in this
kind of models\cite{kessler},
the instantaneous velocity (Fig. \ref{f2} inset) is found to tend to a constant
for large stretching, but it is a fluctuating function of time 
in the low stretching regime.


\begin{figure}
\includegraphics[angle=-90,width=8cm,clip=true]{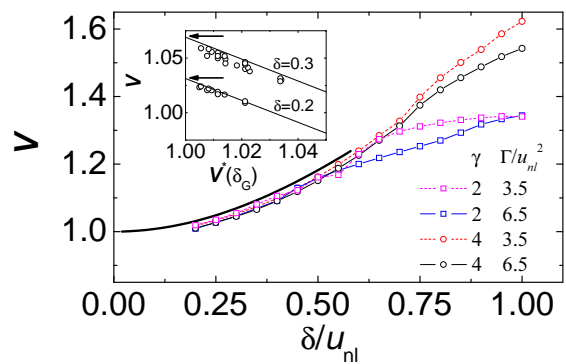}
\caption{(Color online) Velocity of the crack in a system with $N_x=120$, $N_y=240$ for 
different model parameters as a function of stretching.
The thick continuous line is an analytical fitting valid for low $\delta$, large $\gamma$, 
and $N_y\rightarrow \infty$. 
The inset shows that the fitting improves by increasing system size ($N_y\rightarrow \infty$ corresponds 
to $V^*(\delta_G)\rightarrow 1$, the values predicted by the fitting are indicated by the arrows). 
The points reported in the inset correspond 
to all combinations of the parameters: $\gamma=2,4$, $N_y=150,240,360$, and $u_{bk}=1.5,2,2.5,3$, 
plus those in the main plot.
}
\label{f7}
\end{figure}

Results change if $\gamma>1$. 
In this case different velocity curves for different stripe widths do not tend to
collapse in the large $N_y$ limit if plotted
as a function of $\varepsilon$. Instead, they do if plotted as
a function of the unperturbed value $\delta$ of the strain ahead of the
crack. This is an important result: since the propagation is supersonic, the crack 
can respond only to the
local state of the medium, and the velocity must be independent
of the system width if this is large enough.
In Fig. \ref{f7} we plot the velocity dependence on $\delta$ for different system parameters.
We clearly see the supersonic propagation, which occurs for any $\delta>0$ if $N_y$ is large enough.

To analyze these results, it is convenient to solve first the following formal problem.
Consider a perfectly harmonic, infinite and continuous system, stretched some fixed 
amount $\delta$ per unit length in the $y$
direction. 
Suppose that by some external mean we force a crack 
to propagate along the $x$ direction 
at some velocity $V$, which is arbitrary, with the only condition $V>1$. 
The displacement field $u(x,y)$
is found to be simply:

\begin{eqnarray}
u(x,y)&=&y\delta~~~~~~~~~~~~~~~~\mbox{outside the Mach's cone}\nonumber\\
u(x,y)&=& \pm\frac{x\delta}{\sqrt{V^2-1}} ~~~~\mbox{inside the Mach's cone, for }y \lessgtr 0.\label{cono}
\end{eqnarray}
Here, the crack tip is located at $x=0$, $y=0$, and the Mach's cone half-angle $\theta$ is given by $\sin \theta =1/V$.
Note that inside the Mach's cone this solution has a constant value of $|d u/dx|$, 
given by $|d u/dx|=\delta/\sqrt{V^2-1}$. In the case in which the same problem is solved (numerically) on a discrete
square lattice (inducing the propagation by breaking bonds at a fixed rate $V$), 
an oscillation close to the border of the Mach's cone is found, 
and the maximum stretching $|\Delta u|_{\mbox {max}}$ of a spring in the system
is enhanced by a constant factor.
It is found to be 
$|\Delta u|_{\mbox {max}}\simeq 1.26~ \delta/\sqrt{V^2-1}$. Note that $|\Delta u|_{\mbox {max}}$ diverges when
$V\rightarrow 1$ no matter how small $\delta$ is.

\begin{figure}
\includegraphics[angle=-90,width=8cm,clip=true]{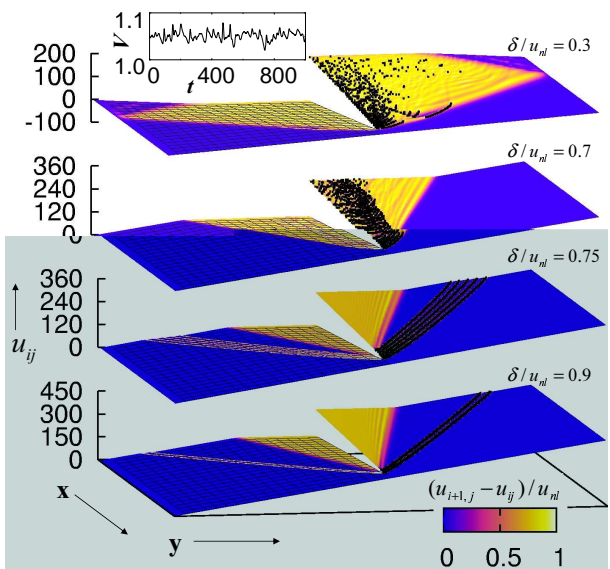}
\caption{(Color online) Four views at different $\delta$ values of the supersonic cracks, for
$\gamma=4$ and $u_{bk}/u_{nl}=2$  in a system of size $120\times 240$.
The $u_{i,j}$ function is plotted (for clarity, we plot the mesh using only 1 every 6 
$\times$ 12 actual points in the left halves of the plots),
and the surface is shadowed according to the values of $(u_{i+1,j}-u_{i,j})/u_{nl}$.
The points of the system at which the non-linear threshold $u_{nl}$ is exceeded
(i.e, when $(u_{i+1,j}-u_{i,j})/u_{nl}>1$)
are highlighted in black on the right halves of the plots. The configuration of the upper 
two panels are not totally stable. 
Instantaneous velocity as a function of time fluctuates as shown for
$\delta/u_{nl}=0.3$, 
and the spatial location of points where the non-linear threshold
is exceeded changes with time.}
\label{f3d1}
\end{figure}

In Fig. \ref{f3d1} we see the actual results from the simulations in the non-linear model.
At low $\delta$ (upper panel) the $u_{i,j}$ function looks roughly similar to that given by Eq. (\ref{cono}) (except
for the reflection effects at the lateral borders of the system).
We find a number of spatial positions 
at which the non-linear threshold is exceeded. This is quite reasonable, otherwise we should not expect
any supersonic propagation. 
In addition, it is found that the instantaneous crack tip velocity is a 
fluctuating function of time, typically around $\pm$
10 \% of the mean value. This fluctuation is acoompanied by a change in the location 
of the points at which the non-linear threshold is exceeded. 
In spite of this, the mean velocity can be determined to a high precision by averaging over a long simulation
time.
Upon increasing $\delta$,
the non-linear regions tend to
arrange in the form of solitons, that eventually (for $\delta/u_{nl}$ between $0.7$ and $0.75$ in Fig. \ref{f3d1})
go outside the Mach's cone. 
There are five solitons at $\delta/u_{nl}=0.75$ in Fig. \ref{f3d1}, and this number reduces as $\delta$ increases.
For the largest $\delta$ values studied, 
typically a single soliton is observed, which drives the supersonic propagation.
We have carefully verified that the solitons in front of the crack
tip are simply the non-linear solitary waves corresponding to the present non linear elastic system, 
i.e., they are the analogous  of the solitons appearing for instance in the  well known Toda 
lattice\cite{toda}. 
In the regime of solitons-driven propagation the
instantaneous velocity is constant.
The transition between the low-$\delta$ and the soliton driven regimes is found to be abrupt, with some hysteresis
upon increasing/decreasing $\delta$, and with a small but clearly observable jump in the velocity (not 
appreciable in the scale of Fig. \ref{f7}).

The low $\delta$ region of the plot in Fig. \ref{f7} shows a remarkable independence on model parameters. Actually,
this part of the curve can be understood and fitted on the basis of the following
heuristic argument. Consider the previously found solution for the harmonic system, Eq.  (\ref{cono}).
We look for the possibility that this solution is self-maintained in the presence of 
non-linearities (i.e., when $\gamma\ne 1$), 
and we analyze in particular the case in which 
$\gamma$ is very large. It is then found that the velocity $V^*$ obtained by requiring 
$u_{nl}=|\Delta u|_{\mbox {max}}$ plays a special role. This velocity is (taking into account the factor introduced
by the discreteness of the system)
\begin{equation}
V^*(\delta)\simeq \sqrt{1+(1.26 ~\delta/u_{nl})^2}.
\label{v}
\end{equation}
In fact, if $V>V^*$, the solution does not explore the non-linear part of the potential anywhere, and then 
it cannot be self-maintained.
On the other hand, if $V<V^*$, there are regions in the system in which the non-linear threshold
is exceeded by a finite amount, but this cannot be acceptable if $\gamma$ is sufficiently large. Then we 
expect that the velocity at which the crack propagation stabilizes is precisely $V^*$. This prediction
is plotted on top of the numerical results in Fig. \ref{f7}. The only parameter of the model
on which $V^*$ depends upon is the non-lineal threshold $u_{nl}$, beyond that the solution is independent
of the precise values of $\gamma$ (assumed large) and $u_{bk}$. 
The fitting improves for larger systems as shown in the inset to Fig. \ref{f7}, where different points for
all combinations of the parameters indicated are plotted as a function of $V^*(\delta_G)$. 
The continuous lines in Fig. \ref{f7} (inset) correspond to the finite size ansatz $V(\delta)=V^*(\delta)-V^*(\delta_G)+1$.
The numerical data follow accurately this trend, and
numerical extrapolation
for $N_y\rightarrow \infty$ allows to claim that the fitting (\ref{v}) is better than 1 \% for $\delta/u_{nl}\lesssim 0.3$
for all the parameters studied. 
This accuracy is remarkably
good because, as we already said, the instantaneous velocity in the low $\delta$ regime fluctuates  around the
mean value in about 10 \%. 

An attempt to predict by the same kind of geometric argument the propagation velocity 
of intersonic mode I or mode II cracks gives negative results:
A theoretical investigation in a purely harmonic system
shows that (contrary to the mode III solution (\ref{cono})) the stress fields 
diverge at the crack tip and are dependent on the boundary conditions
at infinity (see \cite{broberg}, pp. 348-355). Then there is not a well defined maximum 
stress to be used in an argument similar to that presented for mode III.
In addition, the consideration of truly supersonic mode I or mode II cracks for totally harmonic 
springs up to breaking shows that
(in opposition to the mode III case)
for any fixed velocity
all stresses fall below any chosen threshold value if $\delta$ is taken small enough, and this implies 
that supersonic propagation
cannot occur for arbitrarily low values of stretching. 
These considerations show that the present mode 
III problem stands as a very particular but important case in which the crack tip velocity can be predicted on the basis 
of general arguments.

The simulations that produced Fig. \ref{f7} were done by allowing to break only the vertical
springs ahead of the crack. However, we can check a posteriori whether other springs overpass the breaking threshold
$u_{bk}$ or not. It is found in general that there is a separating value $\delta_0$ below which the previous prescription is
not necessary.  If all springs are 
allowed to break, there are no changes in the results for $\delta<\delta_0$, whereas other phenomena (typically
crack branching) are observed if $\delta>\delta_0$. We found $\delta_0$ is
roughly $0.2\sim 0.3$
for the parameters we have simulated. As supersonic propagation occurs in our model for any $\delta$ (in the $N_y\rightarrow
\infty$ limit) we have here an example of supersonic crack propagation in which the crack tip is stable
without the ad hoc introduction of breakable springs located only on a previously defined crack path.

All simulations presented have been done in the absence of any dissipative term. Therefore we can clearly ascribe the
supersonic crack propagation to the non-linearities of the potential, in contrast with other models in which 
supersonic propagation is observed to be induced by the existence of some kind of dissipative terms \cite{marder}.
If in our model a dissipative Kelvin term\cite{kessler} of typical strength $\alpha$ is included 
(i.e., a generic term at the
r.h.s. of Eq. (\ref{eq}) of the form $\sim\alpha \partial(\nabla^2 u)/\partial t $), we have
verified that the behavior of the velocity is smooth when $\alpha \rightarrow 0$, 
re-obtaining the results with no dissipation in this limit. 

It is absolutely clear that the stiffening
of the potential is the responsible for the supersonic crack propagation in this model. 
However, we have observed that the effect of the
stiffening of horizontal and vertical springs on the propagation is very different, and 
in a certain sense, counter intuitive.
Non-linearities of vertical springs do not have a qualitative effect on the results presented: 
If the non-linear threshold for the vertical springs is moved to infinity, the velocity 
curves obtained are only slightly modified, the low $\delta$ fitting (\ref{v}) remains good, and in particular propagation
remains supersonic. On the other hand, if the non-linear threshold of the 
horizontal springs is moved to infinity, supersonic propagation completely disappears.
Then we arrive at the seemingly paradoxical result that the non-linearities of the springs that actually break
are not crucial in determining the propagation velocity. This consideration is 
important for theoretical arguments since
it tells that the present results cannot be explained with an analysis assuming a Barenblatt-type process 
zone (see \cite{broberg}, ch. 3), since in this
case only non-linearities in the vertical links are considered.

In conclusion, we have studied the supersonic propagation of cracks in a lattice model of mode III fracture, in the context
of elastic stiffening at large deformation\cite{gao}. 
Crack velocity is found to depend on the local strain ahead of the crack. For large stretching, the crack propagation is
driven by solitons formed in the non-linear lattice.
In the low stretching regime well developed driving solitons are absent. In this last case
the crack velocity can be accurately predicted on the basis of a geometrical argument, 
and it is found to have a general explicit expression.
This stands as one of very few predictions of crack propagation velocities in models of brittle
fracture. 
 
The authors acknowledge financial support from CONICET (Argentina).

\newpage

\end{document}